\documentclass[prd,preprint,superscriptaddress,longbibliography]{revtex4-2}

\usepackage[utf8]{inputenc}   
\usepackage[T1]{fontenc}
\usepackage{graphicx}
\usepackage{amsmath,amssymb}
\usepackage{float}
\usepackage{url}
\providecommand{\apj}{Astrophys. J.}
\providecommand{\apjl}{Astrophys. J. Lett.}
\providecommand{\apjs}{Astrophys. J. Suppl. Ser.}
\providecommand{\aap}{Astron. Astrophys.}
\providecommand{\mnras}{Mon. Not. R. Astron. Soc.}
\providecommand{\ssr}{Space Sci. Rev.}
\providecommand{\prl}{Phys. Rev. Lett.}
\providecommand{\pre}{Phys. Rev. E}
\providecommand{\jgr}{J. Geophys. Res.}
\providecommand{\grl}{Geophys. Res. Lett.}
\providecommand{\nat}{Nature}

\begin{document}

\title{Microphysical Diversity in Two Very Closely Spaced Magnetic Switchbacks Observed by Parker Solar Probe}

\author{Dipali Vadher}
\affiliation{Department of Physics, Sardar Vallabhbhai National Institute of Technology, Surat, 395007, India}
\affiliation{Space Physics Laboratory, Vikram Sarabhai Space Centre, Thiruvananthapuram, 695022, India}
\email{dipaleevadher09@gmail.com}

\author{Ankush Bhaskar}
\affiliation{Space Physics Laboratory, Vikram Sarabhai Space Centre, Thiruvananthapuram, 695022, India}
\email{ankushbhaskar@gmail.com}

\author{Smitha Thampi}
\affiliation{Space Physics Laboratory, Vikram Sarabhai Space Centre, Thiruvananthapuram, 695022, India}
\email{smitha.v.thampi@gmail.com}

\author{Kamlesh Pathak}
\affiliation{Department of Physics, Sardar Vallabhbhai National Institute of Technology, Surat, 395007, India}
\email{knp@phy.svnit.ac.in}

\begin{abstract}
Parker Solar Probe observations near the Sun reveal frequent, sudden reversals of the magnetic field known as switchbacks (SBs). Despite their ubiquity, the internal plasma structure and associated heating within SBs remain poorly understood. We present a case study of two closely spaced SBs (referred in text as  SB$_1$ and SB$_2$) observed on 24 January 2020 using high-cadence magnetic and plasma measurements. Magnetic fluctuations are decomposed into components parallel and perpendicular to the mean field, and their power spectra are analyzed to characterize the turbulent cascade. The Partial Variance of Increments (PVI) method is applied to identify intermittent current-sheet-like features. Both SB intervals exhibit clear Alfvénic behavior and enhanced radial flow; however, their microphysics differ: SB$_1$ shows a higher proton temperature, larger fluctuation amplitudes, and a denser population of current sheets compared to SB$_2$. The two events also differ in spectral index, with SB$_1$ exhibiting a steeper perpendicular slope than SB$_2$. The elevated intermittency, proton temperature, and transient $\beta>1$ excursion in SB$_1$ suggest that localized dissipation at small-scale structures is a plausible driver of the observed heating. These findings demonstrate that SBs are not uniform kinematic deflections but dynamically evolving plasma structures whose internal turbulence may regulate local energy conversion and contribute to the spatially intermittent heating of the near-Sun solar wind. 
\end{abstract}


%% ---------------------------------------------------------------
%% APS REQUIREMENT FOR arXiv POSTING
%% Once the article is published and the DOI is assigned, replace
%% XXXXXXX below and REMOVE the leading %% from the five lines of
%% \begin{center} ... \end{center} so the notice is typeset.
%%
%% \begin{center}\small
%% Published as Phys. Rev. D \textbf{VV}, NNNNNN (2026),
%% \href{https://doi.org/10.1103/XXXXXXX}{doi:10.1103/XXXXXXX}.\\
%% \copyright{} 2026 American Physical Society. This is the author's
%% accepted manuscript; it is not the version of record.
%% \end{center}
%% ---------------------------------------------------------------

\maketitle

\section{Introduction}

The Parker Solar Probe (PSP) mission~\cite{fox2016} has revealed that magnetic switchbacks (SBs)—abrupt, large-angle kinks in the heliospheric magnetic field—are ubiquitous in the near-Sun solar wind. These events typically involve a reversal of the radial magnetic-field component accompanied by a spike in solar wind speed, and are characterized by nearly Alfvénic fluctuations, where magnetic and velocity perturbations remain closely aligned \cite{Kasper2019, Bale2019, DudokdeWit2020}. Such reversals are often accompanied by enhanced solar-wind flow, reinforcing their strongly Alfvénic nature~\cite{Akhavan-Tafti_2022}.

The origin of these structures remains a key open question. In the solar-origin scenario, SBs are thought to form near the Sun through processes such as interchange magnetic reconnection or jet-like ejections from open magnetic regions \cite{Fisk2020, Raouafi2023, Drake2021}. Recent remote-sensing studies have provided complementary evidence that such reconnection-driven flows in the quiet Sun and coronal holes can give rise to magnetic field kinking and outflows that may evolve into SBs in the nascent solar wind \cite{Upendran2022}. Alternatively, the in-situ hypothesis suggests that SBs can develop locally as the solar wind expands outward, driven by nonlinear Alfvénic turbulence, velocity shear, or large-scale magnetic-field bending \cite{Squire_2020, Tenerani2020, Ruffolo2020}. Determining which of these mechanisms dominates requires detailed case studies that directly connect in-situ plasma measurements with their coronal source environments.  

Large-scale statistical analyses from PSP observations have shown that SB intervals frequently exhibit sharp magnetic gradients and localized discontinuities, indicating the presence of embedded current-sheet-like structures and complex magnetic topology \cite{Huang2023}. These findings suggest that SBs are not merely smooth, large-scale deflections but instead host fine-scale folding and twisting of the magnetic field.  Such small-scale structures have been increasingly associated with plasma heating \cite{Osman2011, Osman2012, Greco2018, Chasapis2018, Wan2016}. Previous studies \cite{Osman2012, Greco2018, Pecora_2022} used the Partial Variance of Increments (PVI) method to show that magnetic intermittency—characterized by localized, bursty magnetic gradients—strongly correlates with elevated proton temperatures in the solar wind. This correlation arises because intermittent regions typically coincide with thin current sheets where magnetic energy is converted into particle kinetic and thermal energy through turbulent dissipation, wave–particle interactions, and reconnection processes \cite{Osman2011, Wan2016, Chasapis2018}. Consequently, the clustering of intermittent structures acts as a natural site of localized plasma heating, linking turbulence-driven intermittency with the observed thermodynamic variability in the solar wind \cite{Osman2012, Pecora_2022}.

Recent PSP observations further indicate that enhanced electron temperatures also coincide with discontinuous magnetic regions \cite{Phillips2023}, suggesting that turbulence-driven reconnection and localized wave activity facilitate inhomogeneous heating across the plasma. This interpretation is supported by kinetic-scale analyses showing that current sheets act as active sites of energy dissipation, directly influencing the solar-wind energy budget \cite{Martinović_2021, Perrone2025}. Together, these results highlight the need for a multi-parameter diagnostic approach that combines magnetic and velocity fluctuations with thermodynamic parameters (such as density, temperature, and plasma $\beta$), and includes electron pitch-angle distributions (PADs) to assess kinetic and energetic signatures of SBs.  

In this study, we analyze two closely spaced SBs (SB$_1$ and SB$_2$) observed on 24 January 2020 during PSP’s fourth solar encounter, listed in the Huang~\textit{et al.} \cite{Huang2023} catalog. These events occur only five minutes apart within nearly identical background solar-wind conditions, providing a rare opportunity to examine how adjacent structures can evolve differently. To quantify these differences, we apply a suite of complementary analyses: field-aligned decomposition of $\delta \mathbf{B}$ \cite{Horbury2008}, power spectral density measurements across the inertial range \cite{Bruno2013}, and the PVI technique to identify current-sheet-like discontinuities \cite{Greco2018}. Electron PADs are also examined to evaluate pitch-angle scattering and anisotropy. To link the in-situ observations with coronal origins, a two-step ballistic backmapping and potential field source surface (PFSS) model~\cite{Macneil2022} is used to trace each SB to its estimated coronal footpoint. 

The study aims to answer three central questions:

(1) Do closely spaced switchbacks exhibit similar or distinct turbulence and heating characteristics?
(2)  Which plasma parameters influence these differences in turbulent evolution and heating efficiency?
(3) How do SBs with contrasting internal structures influence the local solar-wind energetics and dynamics near the Sun?

\section{Data and Methods}

\begin{figure}[htbp!]
  \centering
  \includegraphics[width=0.65\textwidth]{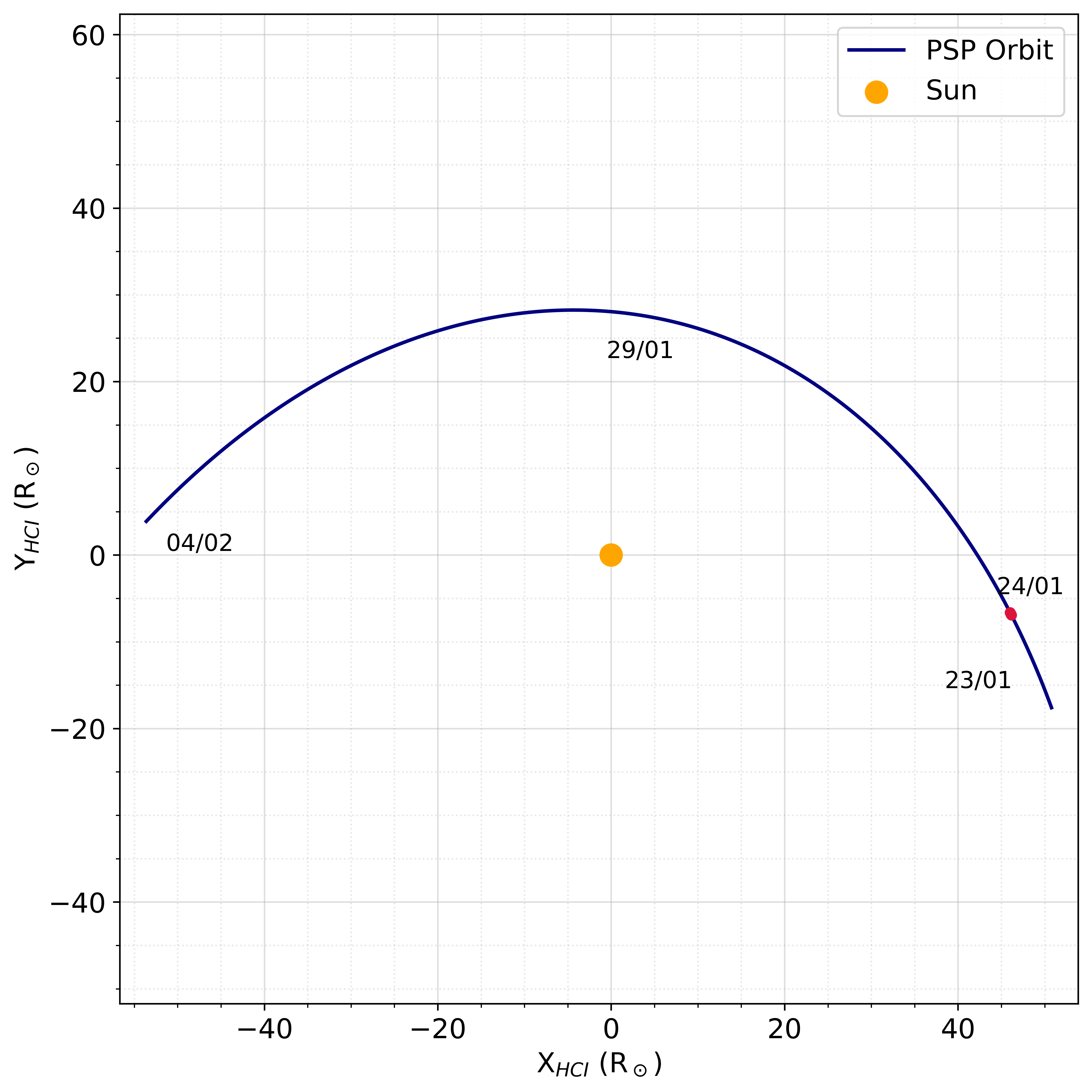}
  \caption{Parker Solar Probe orbit during Encounter 4 (Jan 23 – Feb 4) shown in the Heliocentric Inertial (HCI) frame. The red-highlighted segment marks the interval of interest on 24 Jan 2020, during which SB$_1$ and SB$_2$ were observed.}
  \label{fig:PSP_orbit}
\end{figure}

This study focuses on two consecutively observed magnetic switchbacks (SB$_1$ and SB$_2$) measured by the Parker Solar Probe (PSP) on 24 January 2020 during its fourth solar encounter. The events were identified from the switchback catalog of Huang~\textit{et al.}~\cite{Huang2023}. Figure~\ref{fig:PSP_orbit} illustrates PSP’s trajectory during Encounter 4 (23 January –4 February 2020; perihelion on 29 January at 27.8 $R_\odot$), highlighting the spacecraft’s location on 24 Jan—the interval of interest analyzed in this study.  

The magnetic field vectors in radial–tangential–normal (RTN) coordinates were obtained from the fluxgate magnetometer of the FIELDS instrument suite \citep{Bale2016} at a sampling rate of 4 Hz. Plasma parameters—such as the proton bulk velocity, number density, and thermal speed—were derived from the Solar Probe Cup (SPC), part of the Solar Wind Electrons Alphas and Protons (SWEAP) instrument package \citep{Kasper2016}, which provides measurements at a cadence of approximately 0.874 Hz. To investigate electron-scale behavior, we used pitch-angle distribution (PAD) data from the SPAN-A sensor to characterize the angular spread and anisotropy of suprathermal electrons.

To provide context, the time series from 19:31:00 to 20:18:00~UTC was divided into five intervals: a pre-switchback background interval, Pre-BG (19:31:00–19:41:10), SB$_1$ (19:41:10–19:48:08), Inter-SB (19:48:08–19:53:23), SB$_2$ (19:53:23–20:08:30), and a post-switchback background interval, Post-BG (20:08:30–20:18:00). These boundaries were determined from clear reversals in the radial magnetic-field component ($B_R$) and supported by concurrent variations in velocity, temperature, and plasma~$\beta$. The two background intervals represent relatively undisturbed solar-wind conditions, while the Inter-SB phase captures the short transition between the two switchbacks. SB$_1$ and SB$_2$ correspond to switchbacks No. 23 and No. 24 from the Encounter 4 catalog, both with Quality Flag = 3. The two events are separated by approximately five minutes, which allows a direct comparison of neighboring switchbacks under nearly identical large-scale solar-wind conditions and relative to the same ambient background.

\textit{Field-aligned fluctuation decomposition and spectral analysis}:  
To investigate the geometry and anisotropy of magnetic-field fluctuations, the data were transformed into a field-aligned coordinate (FAC) system following \cite{Horbury2008}. The mean magnetic field vector is defined as
 \(\mathbf{B}_0 = \langle \mathbf{B}(t)\rangle\) defines the local direction. We estimate fluctuations as \(\delta \mathbf{B}(t) = \mathbf{B}(t) - \mathbf{B}_0\), and resolve them into parallel and perpendicular components to mean magnetic field,
\begin{equation}
\delta\mathbf{B}_\parallel = \delta\mathbf{B} \cdot \hat b_0,\quad
\delta\mathbf{B}_\perp = \delta\mathbf{B} - \delta B_\parallel \,\hat b_0,\quad
\hat b_0 = \frac{\mathbf{B}_0}{|\mathbf{B}_0|}.
\label{eq:field_decomp}
\end{equation}

This procedure separates compressive (parallel) fluctuations from transverse Alfvénic (perpendicular) fluctuations. Power spectral densities for both components are estimated using Welch’s method \cite{welch}, which computes the spectrum by dividing the time series into partially overlapping segments, evaluating the periodogram for each segment, and averaging the resulting spectra to reduce variance. In our implementation, the segment length was chosen as $n_{\mathrm{perseg}}=512$. Linear fits in log–log space were performed over the inertial-range band (0.01–0.5 Hz) to obtain the spectral indices \cite{Bruno2013}, and uncertainties were estimated from the standard error of the regression. This approach allows quantitative comparison of turbulence anisotropy and cascade development across intervals.

\textit{Elsässer variables and cross-helicity}: To quantify the Alfvénic content and propagation sense of the fluctuations, we compute the Alfvén speed, the Elsässer fields, and a time-resolved normalized cross-helicity. The Alfvén velocity is
\begin{equation}
\mathbf{V}_A = \frac{\mathbf{B}}{\sqrt{\mu_0\,\rho}},
\label{eq:alfven_velocity}
\end{equation}

where $\rho = n_p m_p$ is the proton mass density and $\mu_0$ is the vacuum permeability. 
From the plasma bulk velocity $\mathbf{V}$ and the Alfvén velocity $\mathbf{V_A}$, we construct the Elsässer variables

\begin{equation}
\mathbf{Z}^{\pm} = \mathbf{V} \pm \mathbf{V_A},
\end{equation}

which represent Alfvénic perturbations propagating parallel and anti-parallel to the local magnetic field \cite{Belcher1971, Bruno2013}. 
In the solar-wind context, these are interpreted as outward (anti-Sunward) $(+)$ and inward (Sunward) $(-)$ propagating fluctuations relative to the Sun. To maintain this interpretation consistently in the heliospheric frame, we apply a polarity correction using the sign of the radial magnetic field component $B_R$, so that $\mathbf{Z}^{+}$ always corresponds to outward (anti-Sunward) propagation and $\mathbf{Z}^{-}$ to inward (Sunward) propagation even during local magnetic field reversals.

The Elsässer fields are complementary quantities that together describe the full Alfvénic dynamics: 
while $\mathbf{Z}^+$ corresponds to fluctuations propagating away from the Sun along the mean magnetic field, 
$\mathbf{Z}^-$ represents those traveling toward the Sun. 
The relative amplitudes of $\mathbf{Z}^+$ and $\mathbf{Z}^-$ therefore indicate the balance between outward and inward propagating wave populations 
and quantify the degree of Alfvénic coupling between velocity and magnetic field fluctuations.

To track rapid changes in this propagation balance, we evaluate the \textit{time-resolved normalized cross-helicity}, defined as

\begin{equation}
\sigma_c(t) = \frac{2 \, \mathbf{V}\cdot\mathbf{V_A}}{|\mathbf{V}|^2 + |\mathbf{V_A}|^2},
\end{equation}

which ranges from $-1$ (purely inward propagation) to $+1$ (purely outward propagation). 
Here positive values of $\sigma_c$ indicate dominance of outward-propagating Alfvénic fluctuations, while negative values indicate inward-propagating fluctuations. 
Physically, $\sigma_c$ measures the instantaneous alignment between plasma velocity and magnetic field fluctuations—hence it directly reflects the dominance of either $\mathbf{Z}^+$ or $\mathbf{Z}^-$. 
Tracking $\sigma_c(t)$ over time allows us to detect sudden reversals or transitions in the direction of Alfvénic energy flow, such as those occurring across SBs.

These diagnostics provide direct measures of the Alfvénic coupling and the directionality of energy propagation across the five intervals analyzed.

\textit{Intermittency and PVI}: In fully developed turbulence, \emph{intermittency} refers to the departure from the global self-similarity predicted by Kolmogorov's K41 phenomenology, manifested as a nonlinear (anomalous) scaling of the structure-function exponents $\zeta(q)$ with order $q$, in place of the linear K41 prediction $\zeta(q)=q/3$, together with a corresponding growth of the kurtosis of the field increments toward smaller scales \cite{Carbone1994,BenziVulpiani2022,Carbone2004}. We do not measure these scaling exponents here. Instead, we use the Partial Variance of Increments (PVI) to locate the sharp, non-Gaussian increments that are the local signature of intermittency in the time series \cite{Osman2012,Greco2018,Pecora_2022}. The PVI is an extension of the Local Intermittency Measure (LIM) introduced by Farge \cite{Farge1992}; it therefore quantifies how strongly an increment departs from a Gaussian distribution rather than identifying a coherent structure directly, and the empirical link between high-PVI events and current sheets has been established through magnetohydrodynamic (MHD) simulations \cite{Greco2008,Greco2018}. For a time lag $\tau=1$~s, the PVI is defined as

\begin{equation} 
\mathrm{PVI}(t,\tau)=\frac{|\mathbf{B}(t+\tau)-\mathbf{B}(t)|}{\sqrt{\langle|\mathbf{B}(t+\tau)-\mathbf{B}(t)|^2\rangle}}, \label{eq:pvi}
\end{equation} 

where the denominator is the root-mean-square (RMS) of magnetic increments computed over a one-hour sliding window. This local normalization allows us to distinguish increments that are unusually large compared to the ambient level of background turbulence. Within this framework, the magnitude of the PVI value provides an operational measure of the strength of a magnetic fluctuation: values around $\mathrm{PVI}\approx 1$ are typical of moderate turbulent variations, values in the range $\mathrm{PVI}\gtrsim 3$ mark strongly non-Gaussian increments, and very large values such as $\mathrm{PVI}>6$ isolate the most abrupt, rare, and intense gradients in the magnetic field; in MHD simulations, events at these amplitudes are preferentially associated with current sheets and other coherent structures \cite{Greco2018,Sioulas2022}. We adopt this empirical association in interpreting high-PVI events below, while emphasizing that the PVI statistic itself measures local non-Gaussianity rather than detecting coherent structures by definition.

Identifying intermittent fluctuations is of interest because these high-amplitude, non-Gaussian increments are, empirically, sites of enhanced nonlinear interactions and energy transfer in the turbulent cascade, and their occurrence and frequency may be linked to processes such as plasma heating and magnetic reconnection in the solar wind \cite{Bruno2013,Greco2012,Sioulas2022}.

\textit{Magnetic connectivity}: The coronal magnetic connection of SB$_1$ and SB$_2$ was estimated using a two-step ballistic–PFSS mapping procedure. The spacecraft’s position at each interval’s midpoint was projected radially outward to the source surface ($R_{ss} = 2.0 R_{\odot}$), assuming a constant solar wind speed of 400 km s$^{-1}$. The longitudinal correction $\Delta \phi = \Omega_{\odot} \Delta t$ accounts for solar rotation, where $\Omega_{\odot}$ is the Carrington rotation rate. These mapped points were used as seeds for PFSS extrapolations to the photosphere, yielding candidate magnetic footpoints. The extrapolations were performed using the open-source \texttt{pfsspy} package \cite{Stansby2020}. 

Previous studies have demonstrated that the uncertainty in ballistic mapping under similar conditions is typically about 10° in longitude \citep{Macneil2022}. Despite this uncertainty, the resulting locations offer a reasonable approximation of PSP’s magnetic connectivity to the coronal source regions associated with the observed SBs.

\section{Observations}

\begin{figure}[htbp!]
\centering
\includegraphics[width=\textwidth]{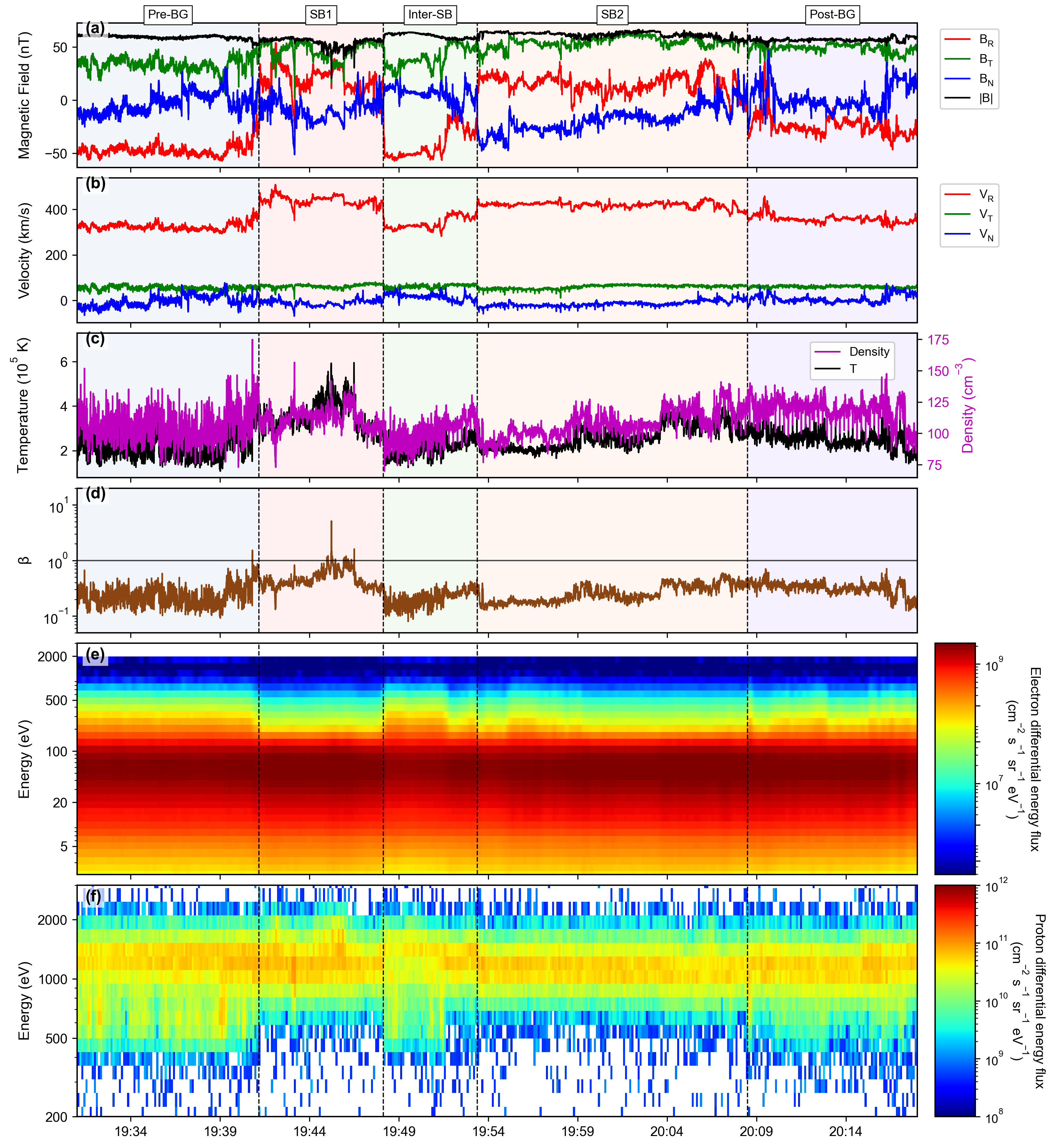}
\caption{Overview of PSP observations on 24 January 2020. 
(a) Radial ($B_R$), tangential ($B_T$), and normal ($B_N$) magnetic field components with total $|B|$. 
(b) Proton bulk velocity in RTN coordinates. 
(c) Proton temperature (black) and number density (magenta).
(d) Plasma $\beta$ (the ratio of thermal to magnetic pressure).
(e) Electron differential energy-flux spectrogram.
(f) Proton differential energy-flux spectrogram. Shaded regions mark the analysis intervals.}
\label{fig:overview}
\end{figure}

Figure~\ref{fig:overview} summarizes the near-Sun plasma and magnetic-field environment recorded by the PSP on 24 Jan 2020. Two consecutive magnetic SBs (SB$_1$ and SB$_2$), stand out against the otherwise steady background flow. Each magnetic deflection is associated with an increase in radial velocity, confirming their strongly Alfvénic character and connecting them to previously reported near-Sun events \cite{Kasper2019,Bale2019,DudokdeWit2020}.

\begin{figure}[htbp!]
  \centering
  \includegraphics[width=0.8\textwidth]{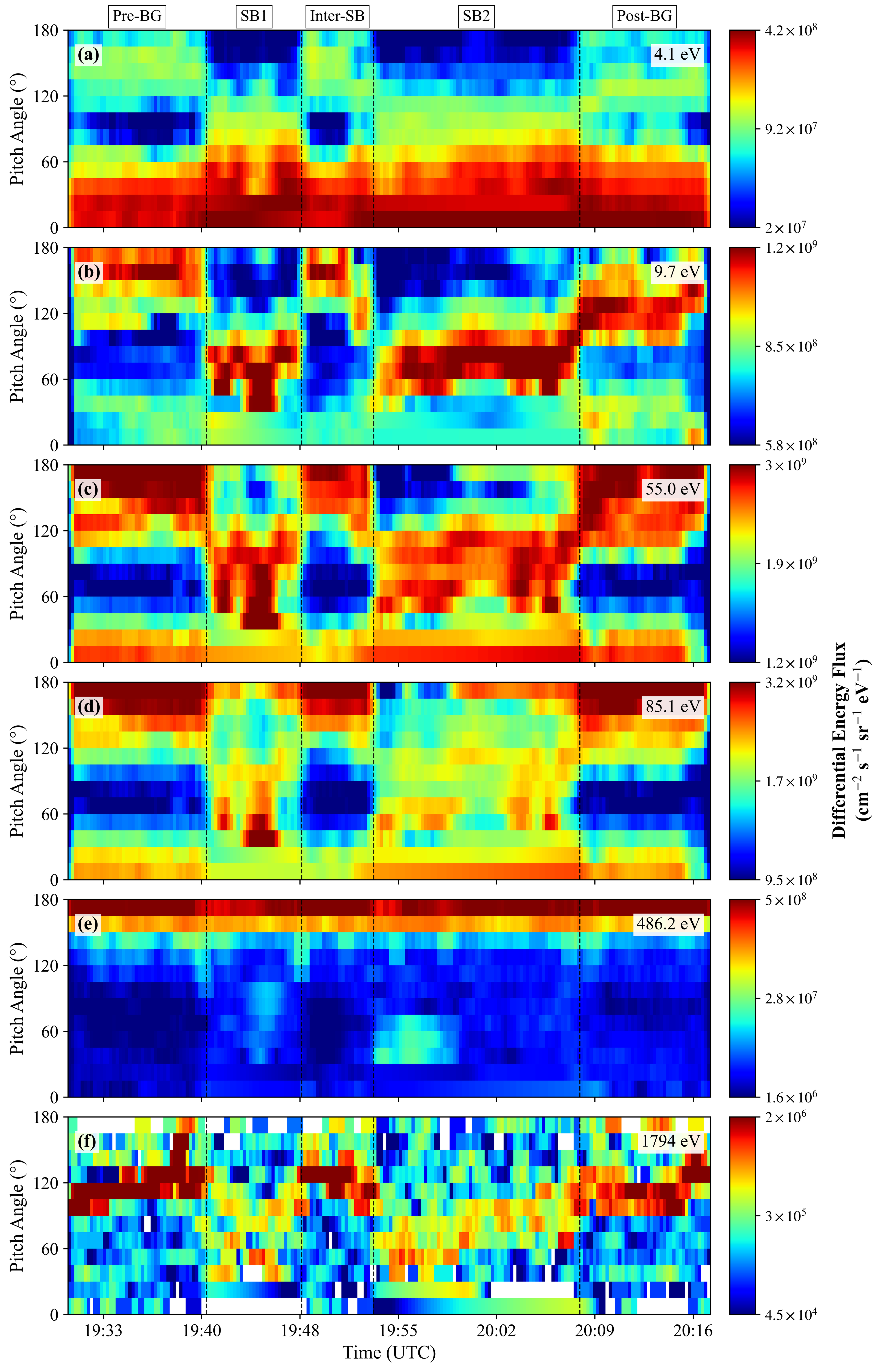}
  \caption[]{Electron pitch-angle distributions (PADs) for various energies across the five intervals: Pre-BG, SB$_1$, Inter-SB, SB$_2$, and Post-BG. The intervals are marked by vertical dashed lines.}
  \label{fig:PADs}
\end{figure}

For detailed comparison, we define five contiguous intervals: Pre-BG, SB$_1$, Inter-SB, SB$_2$, and Post-BG. These segments isolate the two SBs and their surrounding solar-wind conditions for direct contrasts.

During Pre-BG, PSP measured slow-wind conditions with $V\sim330\ \mathrm{km\ s^{-1}}$, $T_p\sim2.1\times10^{5}\ \mathrm{K}$, and plasma~$\beta<1$. SB$_1$ shows a sharp magnetic rotation accompanied by a rapid plasma acceleration; The proton temperature rises to $\sim3.5\times10^{5}\ \mathrm{K}$ and $\beta$ briefly exceeds unity, consistent with localized plasma heating rather than simple adiabatic compression, as reported in previous studies linking enhanced plasma $\beta$ and temperature to dissipation at intermittent structures and reconnection sites \citep{Akhavan-Tafti_2022,Osman2012,Chasapis2018}. The Inter-SB interval marks a partial relaxation of the field and plasma parameters. SB$_2$ presents a second large-angle rotation with an outward velocity spike but a weaker thermal response ($T_p\approx2.5\times10^{5}\ \mathrm{K}$, $\beta<1$). By Post-BG the plasma properties largely return to background values, with only modest residual fluctuations consistent with decaying turbulence.

To investigate the electron response, we examined PADs across multiple energies (Figure~\ref{fig:PADs}). In the background intervals (Pre-BG and Post-BG) the PADs show a narrow anti-sunward \emph{strahl} beam. Within SB$_1$ the strahl broadens markedly across nearly all channels: electrons that were tightly collimated near $0^\circ$ spread toward an approximately isotropic distribution, indicating enhanced pitch-angle scattering likely driven by wave–particle interactions and high-amplitude magnetic fluctuations \cite{Saito2007,Bercic2019}. The Inter-SB interval retains residual broadening, implying continued scattering after the main rotation. In SB$_2$ the strahl refocuses and by Post-BG, returns to the narrow beam typical of slow solar wind.

At lower energies, the background and Inter-SB PADs sometimes show a weaker sunward component near $180^\circ$, consistent with counter-streaming populations along open flux tubes \cite{Horaites2019}. The strongest angular broadening coincides with the highest proton temperatures and turbulence levels (SB$_1$), suggesting that small-scale fluctuations and wave–particle processes redistribute both pitch angle and energy \cite{Vocks2012,Lazar2020}.

\begin{figure}[htbp!]
\centering
\includegraphics[width=0.9\linewidth]{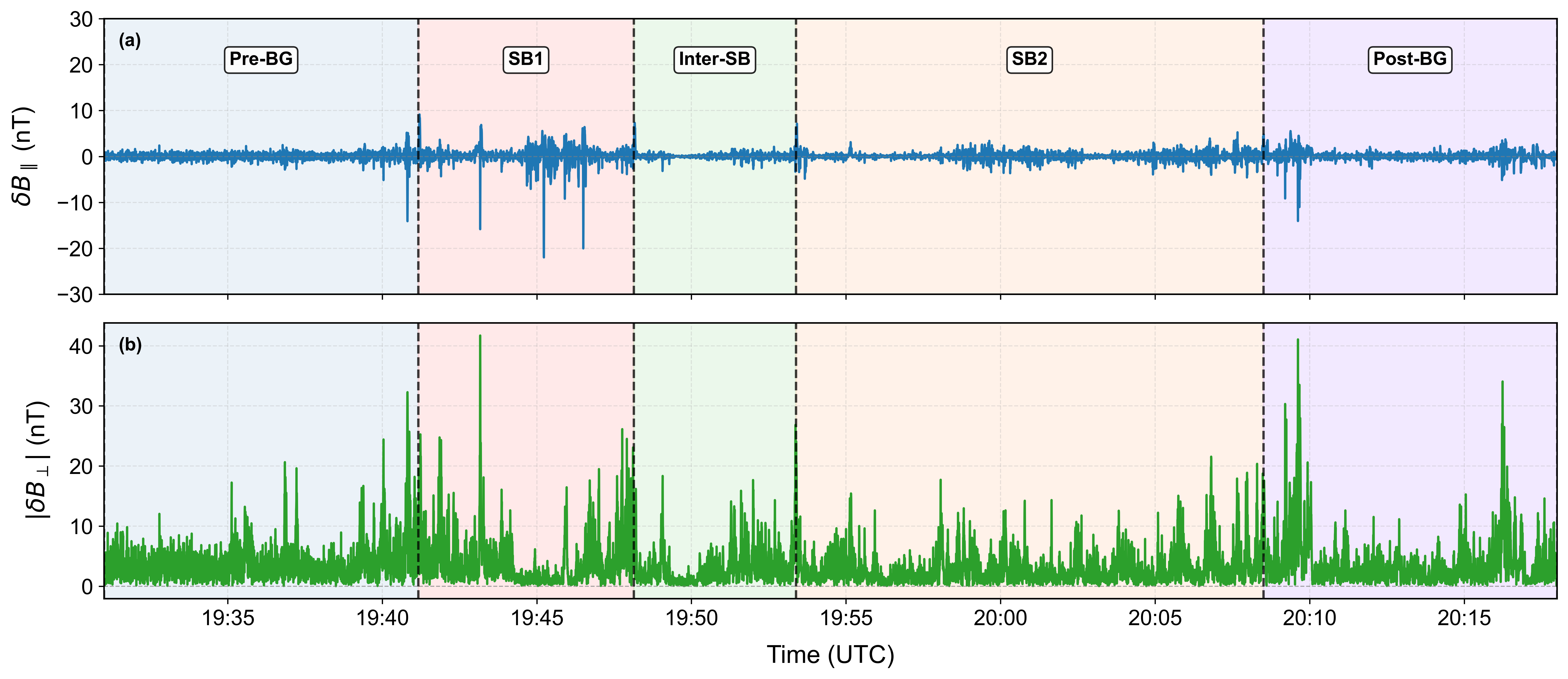}
\caption{Magnetic fluctuations in field-aligned coordinates. (a)parallel component ($\delta B_{\parallel}$).(b) magnitude of perpendicular components ($|\delta B_{\perp}|$).}
\label{fig:FAC}
\end{figure}

Figure~\ref{fig:FAC} shows magnetic fluctuations projected into a field-aligned coordinate system. The perpendicular component $|\delta B_{\perp}|$ dominates the parallel component $\delta B_{\parallel}$ by nearly an order of magnitude in all intervals—behavior typical of near-Sun Alfvénic turbulence \cite{Horbury2008,Belcher1971}. The RMS amplitude of $|\delta B_{\perp}|$ peaks in SB$_1$, decreases in SB$_2$, and is lowest in the background intervals. Occasional spikes in perpendicular power, notably near the start of Post-BG, indicate transient intermittent activity. These trends — enhanced transverse fluctuations in SB$_1$ followed by decay — support the interpretation that SBs transiently amplify local turbulence, which then relaxes as the plasma realigns with the mean field.

\begin{figure}[htbp!]
\centering
\includegraphics[width=\textwidth,height=0.8\textheight,keepaspectratio]{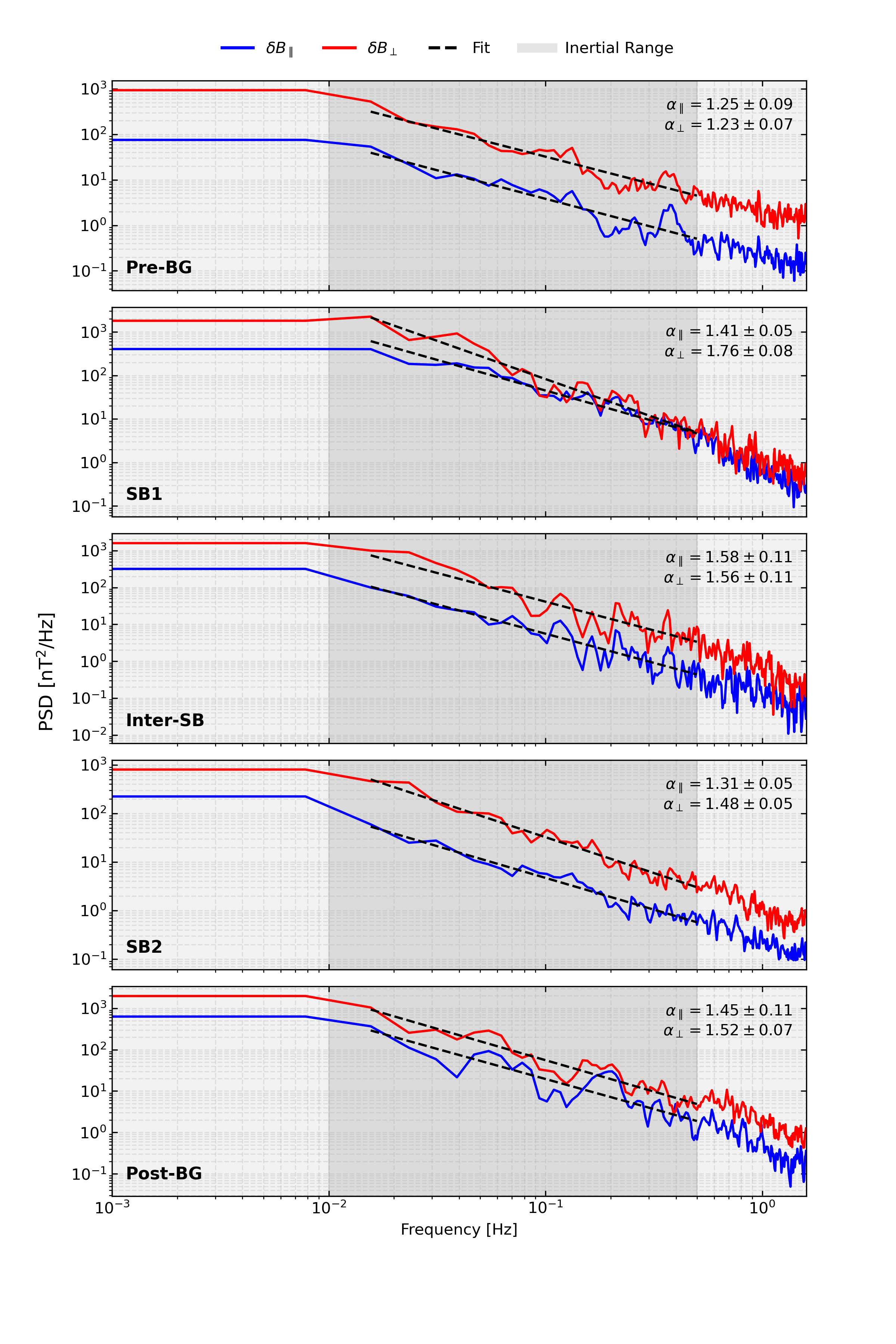}
\caption{Power spectral densities of the field-aligned magnetic fluctuations for the five analysis intervals. Blue curves show the parallel component ($\delta B_{\parallel}$) and red curves show the total perpendicular magnitude ($|\delta\mathbf{B}_{\perp}|$) of variations. Black dashed lines are power-law fits performed in the inertial-range band (shaded gray; 0.01--0.5 Hz). The fitted spectral indices with uncertainties are listed.}
\label{fig:psd} 
\end{figure}

We quantify the turbulent cascade using power spectral densities (PSDs) computed using Welch’s method with a segment length of $n_{\mathrm{perseg}} = 512$ and 50\% overlap (Fig.~\ref{fig:psd}). Power-law fits in the inertial-range band (0.01–0.5~Hz) yield the spectral indices listed in Table~\ref{tab:thermo}. 

The background intervals exhibit relatively shallow slopes ($\alpha_{\parallel}\approx -1.2$ to $-1.5$, $\alpha_{\perp}\approx -1.2$ to $-1.5$), while SB$_1$ shows the steepest slopes ($\alpha_{\parallel}=-1.41\pm0.05$, $\alpha_{\perp}=-1.76\pm0.08$), with the perpendicular component close to the Kolmogorov $-5/3$ value \cite{Kolmogorov1941}. The inter-switchback interval shows intermediate slopes ($\alpha_{\parallel}=-1.58\pm0.11$, $\alpha_{\perp}=-1.56\pm0.11$), and SB$_2$ exhibits comparatively shallower slopes ($\alpha_{\parallel}=-1.31\pm0.05$, $\alpha_{\perp}=-1.48\pm0.05$). These indices are local measurements over short (5.25--15.1~min) windows. Characterizing the turbulent state through the K41 or Iroshnikov--Kraichnan spectral forms requires stationary conditions, that is, averaging over many nonlinear times of the injection scale \citep{Frisch1995}; near perihelion, the outer-scale timescale, defined by the low-frequency spectral break, is itself of order $T\approx8$~minutes \citep{Perez2021}, so several of our analysis windows span at most a few such times. We therefore do not interpret the slope differences across intervals as evidence of a more or less developed cascade, and report them only as local spectral properties of each interval. The steeper perpendicular spectrum in SB$_1$ coincides with its elevated velocity, proton temperature, and density, which remain the direct evidence for enhanced energy transfer there \citep{Howes2008,Verdini2018}.

\begin{table}[htbp!]
\caption{Plasma and spectral properties. Median and maximum values are shown for velocity ($V$), temperature ($T$), and density ($n$).\label{tab:thermo}}
\begin{ruledtabular}
\begin{tabular}{lcccccccc}
Interval & $V_{\mathrm{med}}$ & $V_{\max}$ & $T_{\mathrm{med}}$ & $T_{\max}$ & $n_{\mathrm{med}}$ & $n_{\max}$ & $\alpha_{\parallel}$ & $\alpha_{\perp}$ \\
 & (km s$^{-1}$) & (km s$^{-1}$) & (K) & (K) & (cm$^{-3}$) & (cm$^{-3}$) & & \\
\hline
Pre-BG   & 327.2 & 410.5 & $2.1 \times 10^{5}$ & $7.0 \times 10^{5}$ & 104.7 & 174.7 & $-1.25 \pm 0.09$ & $-1.23 \pm 0.07$ \\
SB$_1$   & 443.7 & 509.9 & $3.5 \times 10^{5}$ & $5.9 \times 10^{5}$ & 112.3 & 156.5 & $-1.41 \pm 0.05$ & $-1.76 \pm 0.08$ \\
Inter-SB & 334.5 & 445.9 & $2.1 \times 10^{5}$ & $4.0 \times 10^{5}$ &  99.8 & 123.2 & $-1.58 \pm 0.11$ & $-1.56 \pm 0.11$ \\
SB$_2$   & 426.1 & 457.5 & $2.5 \times 10^{5}$ & $4.0 \times 10^{5}$ & 105.9 & 141.0 & $-1.31 \pm 0.05$ & $-1.48 \pm 0.05$ \\
Post-BG  & 362.3 & 461.6 & $2.5 \times 10^{5}$ & $4.1 \times 10^{5}$ & 120.1 & 147.5 & $-1.45 \pm 0.11$ & $-1.52 \pm 0.07$ \\
\end{tabular}
\end{ruledtabular}
\end{table}

\begin{figure}[htbp!]
\centering
\includegraphics[width=1\linewidth]{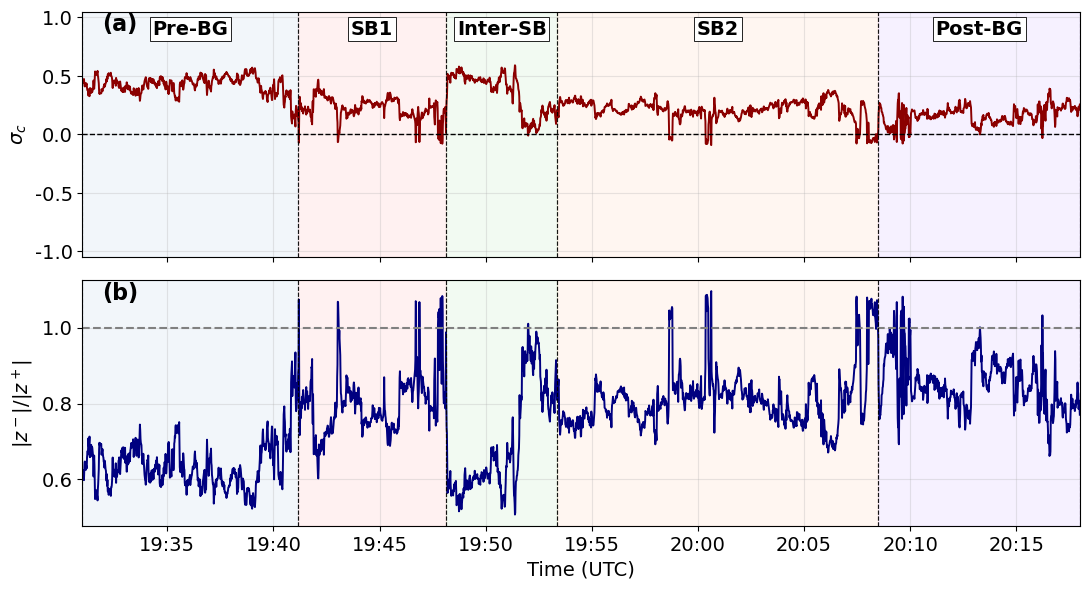}
\caption{(a) normalized cross helicity ($\sigma_c$) and (b) Elsässer ratio ($|z^-|/|z^+|$). The dashed lines mark $\sigma_c=0$ in panel (a) and $|z^-|/|z^+|=1$ in panel (b).}
\label{fig:Alfvenicity}
\end{figure}

To examine Alfvénicity, we computed the Elsässer variables and the normalized cross-helicity; the time series are shown in Figure~\ref{fig:Alfvenicity}. The Pre-BG interval is characterized by moderately positive values of $\sigma_c$ and $|z^-|/|z^+| < 1$, indicating a dominance of outward-propagating Alfvénic fluctuations typical of the ambient solar wind \cite{Matthaeus1982,TuMarsch}. Entering SB$_1$, $\sigma_c$ decreases and the Elsässer ratio approaches unity, suggesting a reduction of the Alfvénic imbalance and a more mixed propagation state. 
During the Inter-SB interval, $\sigma_c$ temporarily increases again, indicating a brief return to stronger outward Alfvénic dominance. In SB$_2$, $\sigma_c$ remains positive but with moderate fluctuations, while $|z^-|/|z^+|$ stays close to unity, implying the coexistence of outward and inward propagating components. Finally, in Post-BG, $\sigma_c$ gradually decreases as the turbulence relaxes. Overall, although outward-propagating Alfvénic fluctuations dominate throughout the interval, the switchback regions exhibit a reduced imbalance and enhanced interaction between counter-propagating modes.

\begin{figure}[htbp!]
\centering
\includegraphics[width=1\linewidth]{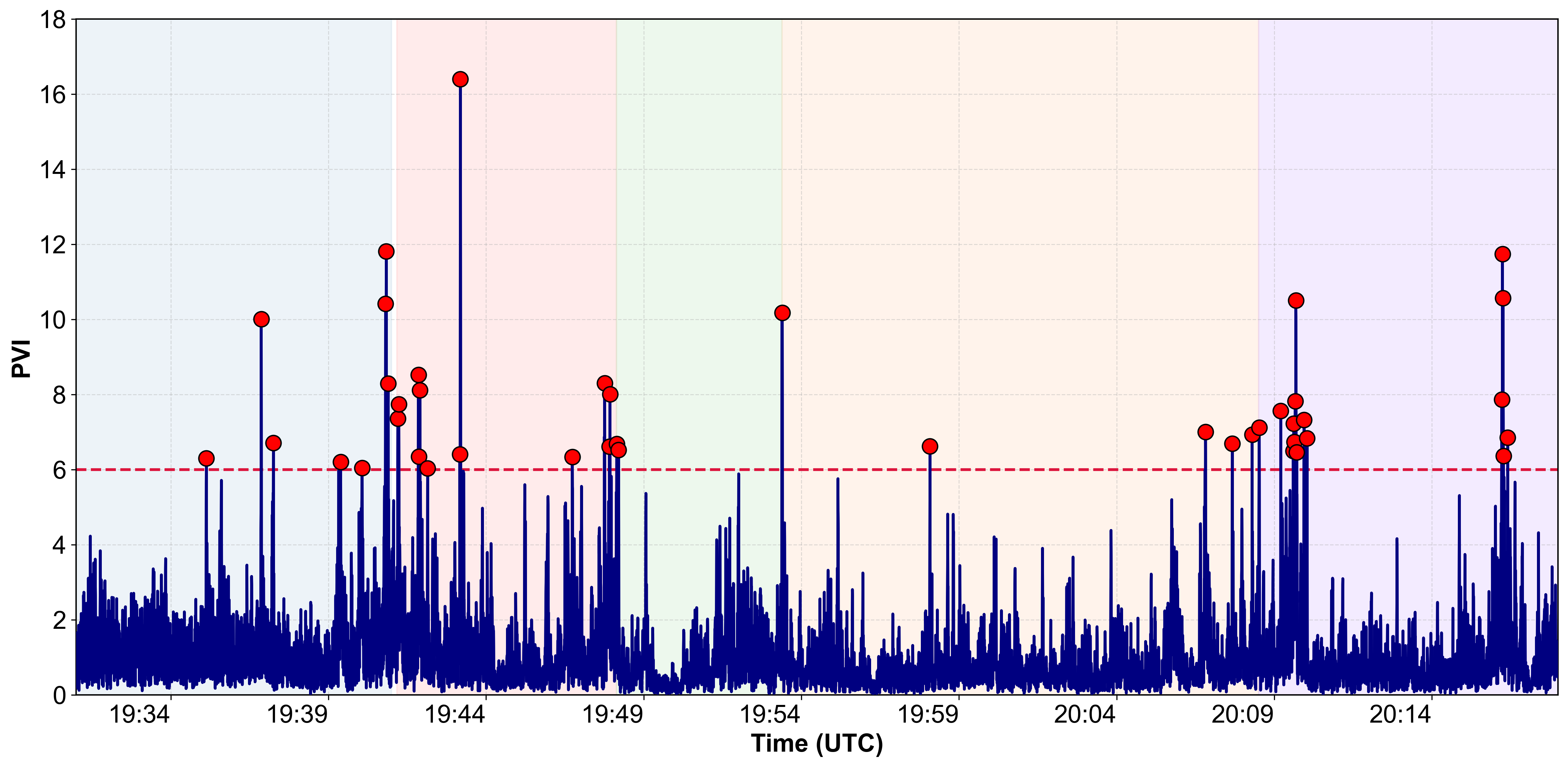}
\caption{Partial Variance of Increments (PVI) time series. The dashed line marks the threshold $\mathrm{PVI} = 6$. Red markers denote current-sheet candidates.}
\label{fig:PVI}
\end{figure}

Small-scale intermittency was quantified with the PVI statistic (Figure~\ref{fig:PVI}). Using $\tau=1$~s and the one-hour RMS normalization, events with PVI$>6$ were treated as strong discontinuities/current-sheet candidates \cite{Osman2012,Greco2018,Pecora_2022}. We find eight such events (0.8~min$^{-1}$) in Pre-BG, twelve (1.72~min$^{-1}$) in SB$_1$, two (0.38~min$^{-1}$) in Inter-SB, five (0.33~min$^{-1}$) in SB$_2$, and fifteen (1.58~min$^{-1}$) in Post-BG. The rates in SB$_1$ and SB$_2$ carry Poisson counting uncertainties of $1.72\pm0.50$~min$^{-1}$ and $0.33\pm0.15$~min$^{-1}$, respectively. Two simple null hypotheses fail to account for this contrast. First, if strong discontinuities occurred at a uniform rate, the longer SB$_2$ interval (15.1~min vs.\ 6.97~min) would contain more events than SB$_1$, whereas it contains fewer (five versus twelve). Second, if high-PVI events arose solely from the interface current sheets expected at the boundaries between counter-oriented magnetic fields, each switchback would contribute only the small number of events set by its two boundary crossings; the twelve events within SB$_1$ far exceed this expectation. The elevated PVI activity in SB$_1$ therefore reflects enhanced small-scale intermittency within the structure rather than an artifact of interval duration or of the boundary reversals alone. We caution that a high rate of strong discontinuities is not by itself a heating diagnostic---the Post-BG interval shows an elevated rate (1.58~min$^{-1}$) without a comparable temperature enhancement---and the case for enhanced dissipation in SB$_1$ rests on the coincidence of its high event rate with the strongest PAD broadening, elevated proton temperature, and the transient $\beta>1$ excursion, suggesting that current-sheet formation and localized reconnection contributed to the enhanced heating there \cite{Greco2018,Chasapis2018,Osman2012}.

\begin{figure}[htbp!]
    \centering
    \includegraphics[width=0.95\textwidth]{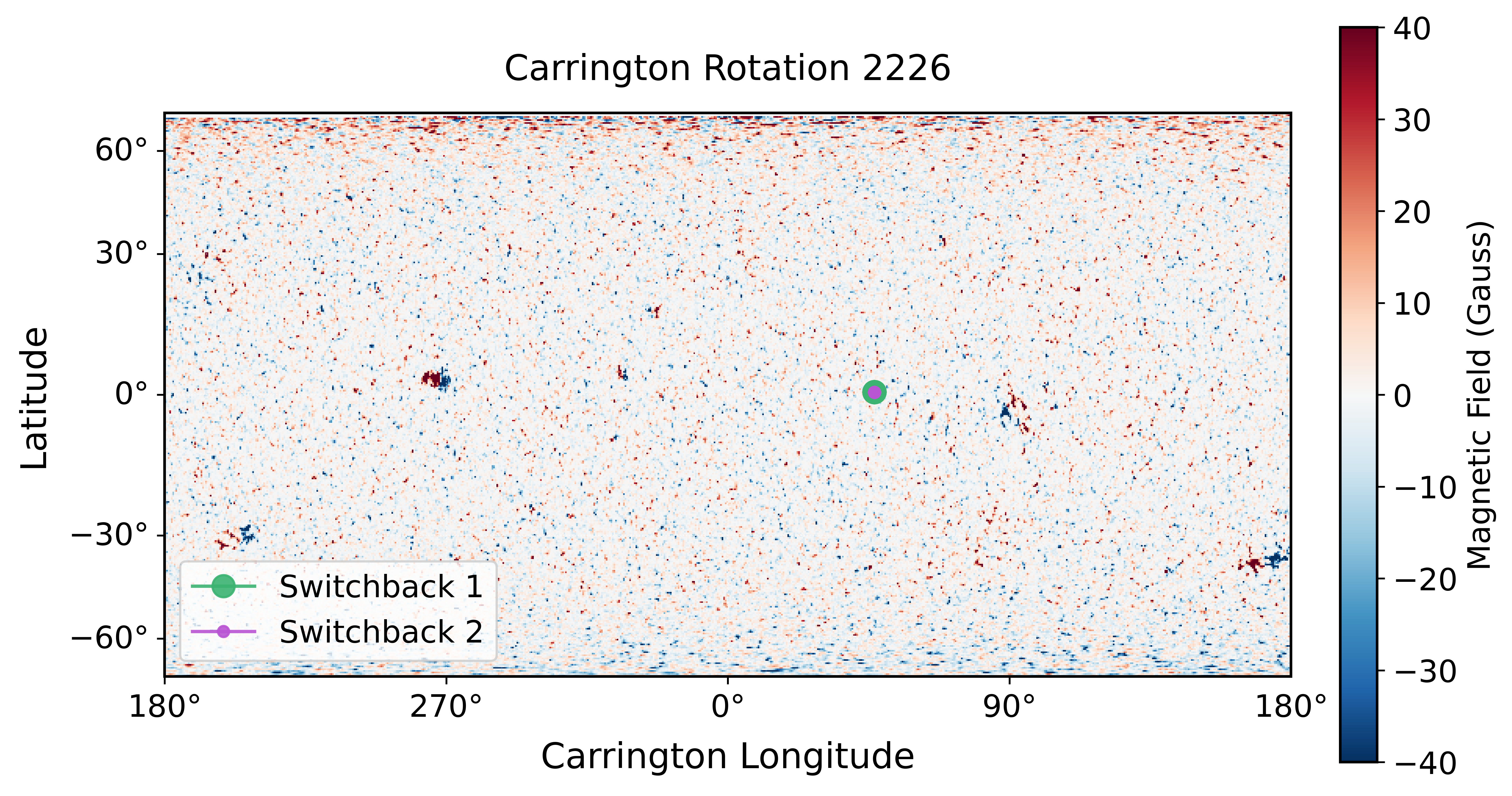}
    \caption{
        PFSS extrapolation of the coronal magnetic field for Carrington Rotation~2226, based on the HMI synoptic magnetogram.
        Red and blue shading denote outward and inward radial magnetic fields, respectively.
        The ballistic mapping results for SB$_1$~({green} markers) and SB$_2$~({magenta} markers) are overplotted at the source surface.}
    \label{fig:pfss}
\end{figure}

Figure~\ref{fig:pfss} shows a PFSS extrapolation for Carrington Rotation~2226 based on the Helioseismic and Magnetic Imager (HMI) synoptic map. We performed two-step ballistic backmapping followed by PFSS extrapolation to estimate coronal footpoints \cite{Macneil2022}. The inferred footpoints for SB$_1$ (46.82° lon, 0.52° lat) and SB$_2$ (46.79° lon, 0.51° lat) lie within the same low-latitude open-flux region, suggesting a common coronal domain likely favorable for interchange reconnection. Although ballistic mapping typically carries a longitudinal uncertainty of order $10^\circ$, the close proximity of the two footpoints supports a shared magnetic origin \cite{Macneil2022}.

Although SB$_1$ and SB$_2$ map to nearly the same coronal region, their in-situ properties differ markedly: SB$_1$ is hotter, more intermittent, and exhibits stronger perpendicular fluctuations. These contrasts imply short-timescale variability in the release or propagation of energy from the coronal source, producing adjacent structures with distinct microphysical states.

\section{Discussion \& Summary}

The analysis of the two closely spaced SBs (SB$_1$ and SB$_2$) shows that even neighboring magnetic structures can differ sharply in their plasma and turbulence characteristics despite occurring under nearly identical solar-wind conditions. Both SBs display the classic Alfvénic signatures—enhanced outward flow and magnetic-field rotation, but their internal microphysics and heating responses are not the same.

\begin{figure}[htbp!]
    \centering
    \includegraphics[width=\textwidth]{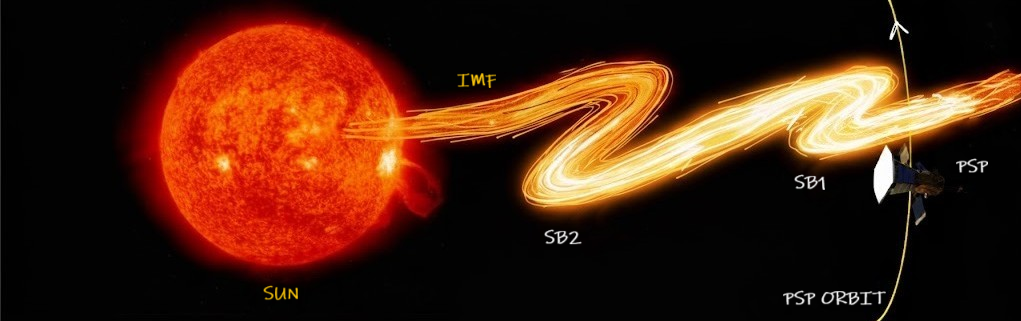}
    \caption{ Artistic illustration showing the Parker Solar Probe (PSP) encountering two magnetic switchbacks (SB$_1$ and SB$_2$). 
The brightness variation along the magnetic field line represents the local plasma temperature: bright yellow regions correspond to enhanced heating within SB$_1$, while the darker red tones in SB$_2$ indicate comparatively lower temperatures.}
    \label{fig:SB_SKETCH}
\end{figure}

SB$_1$ stands out by its higher proton temperature, steeper spectral slopes, and a much larger number of strong intermittent events than SB$_2$. The rate of intermittent events in SB$_1$ (1.72~min$^{-1}$) is over five times that in SB$_2$ (0.33~min$^{-1}$), indicating a stronger presence of localized current-sheet-like structures. These intermittent structures correspond to sharp magnetic gradients where strong shears and compressions trigger enhanced dissipation through reconnection and kinetic processes. The clustering of such discontinuities therefore points to sites of active small-scale energy conversion, consistent with earlier findings that intermittent magnetic structures act as localized dissipation channels linked to plasma heating in the solar wind \cite{Osman2012, Greco2018, Wan2016, Chasapis2018}. Figure~\ref{fig:SB_SKETCH} illustrates this contrast schematically.

The spectral slopes further illustrate these differences: SB$_1$ exhibits a steep perpendicular spectrum with $\alpha_{\perp}\approx-1.76$, close to the Kolmogorov $-5/3$ value \citep{Kolmogorov1941}, whereas SB$_2$ shows a comparatively shallower slope of $\alpha_{\perp}\approx-1.48$. As discussed in Sec.~III, our analysis windows do not satisfy the stationarity condition required to read this contrast as evidence of a more or less developed turbulent cascade \citep{Frisch1995,Perez2021}; we therefore report the slope difference only as a measured property of each interval. The direct evidence for enhanced dissipation in SB$_1$ is instead its elevated proton temperature and a transient period of $\beta>1$, consistent with heating driven by turbulent dissipation rather than adiabatic compression alone \citep{Osman2011,Pecora_2022}.

Cross-helicity and Elsässer diagnostics show that both SBs remain predominantly Alfvénic, with fluctuations dominated by outward (anti-Sunward) propagation relative to the Sun. However, neither interval reaches a purely Alfvénic state ($|\sigma_c|<1$), indicating that both outward and inward propagating components coexist. The presence of counter-propagating fluctuations suggests nonlinear interactions that sustain the turbulent cascade \citep{Matthaeus1982,TuMarsch}. The stronger imbalance and larger fluctuation amplitudes observed in SB$_1$ are consistent with enhanced nonlinear coupling and higher dissipation efficiency compared to SB$_2$.

Ballistic mapping combined with PFSS extrapolation reveals that SB$_1$ and SB$_2$ trace back to nearly the same low-latitude open magnetic region on the Sun (Figure~\ref{fig:pfss}). Their footpoints lie within a common magnetic domain, possibly implying that the differences seen in observations arise from temporal variability in coronal energy release rather than spatial separation of sources. Similar variability has been reported in cases where successive reconnection jets or flux-rope ejections generate SBs of different strength \citep{Hou2024,Drake2021}.

Overall, these observations indicate that SBs are not static field rotations but dynamically evolving Alfvénic plasma structures, whose internal intermittency is associated with the effectiveness of magnetic energy conversion into heat. The enhanced PVI activity within SB$_1$ coincides with stronger heating, consistent with a link between intermittency, dissipation, and thermal response; the spectral contrast between the two intervals is reported separately and is not used here as independent evidence for that link.

This study reveals that even neighboring SBs can differ markedly in their internal turbulence and dissipation efficiency. Differences in intermittency and Alfvénic coupling are associated with how effectively each structure converts magnetic energy into heat, suggesting that microphysics shapes the energetic role of SBs in the solar wind. SBs thus emerge as dynamic agents of intermittent heating, central to the small-scale energy budget of the near-Sun solar wind.

\begin{acknowledgments}
We thank the NASA \textit{Parker Solar Probe} mission and instrument teams for providing the data used in this study. Magnetic field and plasma measurements were obtained from the FIELDS and SWEAP instruments, accessed via NASA’s CDAWeb.
D.~V. acknowledges an Institute Fellowship from Sardar Vallabhbhai National Institute of Technology, Surat, awarded on the basis of the GATE qualification under Ministry of Education, Government of India norms, and thanks SVNIT for the research facilities provided. D.~V. also thanks the Space Physics Laboratory, Vikram Sarabhai Space Centre, for hosting the internship during which this work was carried out.

\end{acknowledgments}

\section*{Data Availability}

The Parker Solar Probe data analyzed in this study are publicly available from NASA's Space Physics Data Facility via CDAWeb: the FIELDS fluxgate magnetometer Level~2 magnetic field vectors in RTN coordinates~\cite{data:PSPMAG}, the SWEAP/SPC Level~3 proton moments~\cite{data:PSPSPC}, and the SWEAP/SPAN-A Level~3 electron pitch-angle distributions~\cite{data:PSPSPAN}.

The Parker Solar Probe Encounter~4 switchback catalog compiled by Huang \textit{et al.}~\cite{Huang2023}, from which SB$_1$ and SB$_2$ were identified, is available at
\url{http://sprg.ssl.berkeley.edu/data/psp/data/sci/sweap/lists/Switchbacks_PSP/}.

The photospheric synoptic magnetogram used for the potential-field source-surface extrapolation is the SDO/HMI line-of-sight synoptic map for Carrington Rotation~2226~\cite{data:HMI}. The extrapolation was performed with the open-source \texttt{pfsspy} package~\cite{Stansby2020}.

\bibliography{apssamp}

\end{document}